# Rigorous determination of the ground-state phases and thermodynamics in an Ising-type multiferroic $Ca_3CoMnO_6$ chain

**Yan Qi[a], Qi Yang[a], Nai-sen Yu[a,*] and An Du[b]**

[a] School of Physics and Materials Engineering, Dalian Nationalities University, Dalian 116600, People's Republic of China

[b] College of Sciences, Northeastern University, Shenyang 110819, People's Republic of China

**Abstract:** To understand the collinear-magnetism-driven ferroelectricity in multiferroic $Ca_3CoMnO_6$ compound, we have established an elastic diatomic Ising spin-chain model with axial-next-nearest-neighbor interaction to describe its magnetoelectric properties. By employing magneto-phonon decoupling and transfer-matrix method, the possible ground-state configurations and thermodynamic behaviors of the system have been exactly determined. From the perspective of the ground-state configuration, we analyze the computational results and make a detail comparison with experimental data. The parameter relation for the appearance of electric polarization has been discussed. Our data indicate that the magnetic coupling between nearest-neighbor spin pair is antiferromagnetic rather than ferromagnetic. The system under the driven of external magnetic field undergoes a different series of transitions from the ↑↑↓↓ spin configuration to the ↑↓↑ state with peculiar 1/3 magnetization plateau, then to the ↑↑↑↓ state, and finally saturated at ↑↑↑↑ state.

## I. Introduction

Multiferroics simultaneously exhibit several ferroic orders such as (anti)ferromagnetism, ferroelectricity, and ferroelasticity. The coupling between the order parameters brings about unusual physical properties and provides possibilities for new multifunctional device applications.[1,2] In recent years, the magnetically driven multiferroics has attracted remarkable attention because of their prominent magnetoelectric (ME) effects.[3] Three major microscopic theories have been proposed to illustrate the coupling between magnetic and ferroelectric orders. One is based on exchange-striction mechanism through the symmetric spin exchange interaction; the other two deal with spin-current model and spin dependent p-d hybridization respectively. First-principle calculations predict that the ferroelectric polarization arising from exchange-striction mechanism can arrive at a considerable value.[4,5] Unfortunately, the actual results are barely satisfied. The underlying microscopic mechanism is still a challenging task and needs further investigation.

Ising chain magnet $Ca_3CoMnO_6$ is a typical collinear multiferroic with its ferroelectricity originating from exchange striction. It is composed of parallel one-dimensional $CoMnO_6$ chains. Each $CoMnO_6$ chain consists of face-shared $CoMnO_6$ trigonal prisms and $MnO_6$ octahedra alternatively along the $c$ axis. These chains are separated by non-magnetic Ca ions and form a triangular lattice in the $ab$ plane.[6,7] Neutron diffraction shows that $Ca_3CoMnO_6$ exhibits a special ↑↑↓↓ magnetic order below temperature $T_N \approx 13K$ and magnetic field $H < 10$ T.

Combined with the alternating $Co^{2+}$ and $Mn^{4+}$ ionic order, this antiferromagnetic structure breaks the lattice inversion symmetry, and leads to the development of ferroelectricity through exchange striction.[8-11]

Ca$_3$CoMnO$_6$ has much stronger intrachain interactions than the interchain ones and Ising-like anisotropy, therefore it can be characterized by an elastic Ising-spin chain model with the competing nearest-neighbor (NN) and next-nearest-neighbor (NNN) interactions (ANNNI model).[12] Several groups based on this model have investigated the fascinating magnetoelectric properties in Ca$_3$CoMnO$_6$. Yao et al. reproduced the low-temperature ferroelectric and magnetic behaviors, and they further explored the responses of the microscopic domain structures of spins and ionic displacements to the external magnetic field, confirming the ferroelectricity in this compound driven by the ↑↑↓↓ collinear magnetic order.[13,14] Guo et al. constructed the mean-field theory to analyze the generation of ferroelectricity by taking into account the intrachain interactions and dynamics of polarization domains.[15] They have found that the stability of the short-range spin order determined the ferroelectric transition. Nishida et al. using the Green's function method evaluated the exchange coupling constants, and they clarified the effects of intra-chain and inter-chain exchange interactions on magnetic stability.[16] Although the research on the magnetoelectric behaviors in Ca$_3$CoMnO$_6$ have achieved some progress, controversies upon magnetic orders of $Co^{2+}$ and $Mn^{4+}$ ions remain unsolved, and the results from these research are not fully consistent with each other.[17-20] In this paper, we make an attempt to illustrate this discrepancy from the perspective of its ground-state orders analysis, for which a simple and effective elastic ANNNI chain model will be used. By using Morita-Horiguchi technique[21], rigorous determinations on the ground-state spin configurations are carried out to describe ferroelectricity in Ca$_3$CoMnO$_6$. And as far as we know, many reports focus on its thermodynamics, but little knowledge is known about ground state properties for this elastic ANNNI chain magnet. Subsequently, we use the transfer-matrix method for exactly calculating the magnetoelectric properties at finite temperature to further elaborate microscopic ME mechanism in this one-dimensional system. Based on the exact solutions, the complex interplay between the physics of frustration, external fields and magnetoelectric effect in this system is explored. We believe that present work will help understanding exotic magnetoelectric behaviors in Ca$_3$CoMnO$_6$ compound, and shed some light on physics of frustrated Ising chains.

## II. Model and Method

Considering the ↑↑↓↓ spin order mainly determined by the intrachain interaction, the diatomic Ising-chain model is introduced to describe the magnetism-driven ferroelectricity in this quasi-one-dimensional system. We assume the NN elastic interactions and NNN AFM interactions, and the Hamiltonian for the model is written as

$$\mathcal{H} = -J_1^{\text{Co-Mn}}\left(r_{i,i+1}\right)\sum_{i=1}^{N}\left(S_{2i-1}+S_{2i+1}\right)S_{2i} - J_2^{\text{Co-Co}}\sum_{i=1}^{N}S_{2i-1}S_{2i+1} - J_3^{\text{Mn-Mn}}\sum_{i=1}^{N}S_{2i}S_{2i+2}$$

$$-H\sum_{i=1}^{N}\left(g_{\text{Co}}\mu_B S_{2i-1}+g_{\text{Mn}}\mu_B S_{2i}\right) - E\sum_{i=1}^{N}q\left(d_{2i-1}+d_{2i}\right) + \frac{1}{2}k\sum_{i=1}^{N}\left(d_{2i-1}^2+d_{2i}^2\right)$$

(1)

Here $S_i = \pm 1$ represents the spin at the $i$th site of the chain. For clarity, it is assumed that the odd sites are occupied by Co ions and even sites are occupied by Mn ions. $J_1^{\text{Co-Mn}}$ denotes the NN elastic coupling between Co ion and Mn ion. $J_2^{\text{Co-Co}}$ ($J_3^{\text{Mn-Mn}}$) refers to the interaction between two NNN Co (Mn) ions. Therefore, the first three terms of Eq. (1) represent the exchange energy between the NN and NNN spin pairs. The fourth and fifth terms correspond to the magnetic field energy and electric field energy, with the symbols of $H$ and $E$ denoted to the external magnetic field and electric field. $g_{\text{Co}}$ and $g_{\text{Mn}}$ are the Lande factors of Co ions and Mn ions, and $\mu_B$ is the Bohr magnon. $q_{\text{Co}}$ and $q_{\text{Mn}}$ are the electric charges. The last term is the elastic energy, where $k_{\text{Co}}$ and $k_{\text{Mn}}$ are the elastic constants. It is assumed that these two kinds of spins share the same magnitude of the following quantities $g_{\text{Co}} = g_{\text{Mn}} = g$, $q_{\text{Co}} = q_{\text{Mn}} = q$ and $k_{\text{Co}} = k_{\text{Mn}} = k$.

The elastic exchange coupling is only applied to the NN spin pair since it cannot give rise to effective exchange striction for the NNN spin pair. Under the linear approximation, $J_1^{\text{Co-Mn}}$ is expanded to the following form [13,14]

$$J_1^{\text{Co-Mn}} = J_0^{\text{Co-Mn}}\left(1+\eta\frac{r_{2i-1,2i}-r_0}{r_0}\right)$$

$$= J_0^{\text{Co-Mn}}\left[1+\eta\left(d_{2i}-d_{2i-1}\right)\right]$$

(2)

Here $J_0^{\text{Co-Mn}}$ and $r_0$ respectively denote the interaction and original distance between NN pair in the absence of exchange striction. $\eta$ describes the strength of coupling between the spins and displacements. For convenience, $\eta = -1$ is chosen here, indicating the strengthen of the coupling for the neighboring magnetic ions come closer to each other. $d_{2i-1}$ ($d_{2i}$) denotes the reduced displacements of Co (Mn) ions, and their movements are determined by the coupling type of $J_0^{\text{Co-Mn}}$. When $J_0^{\text{Co-Mn}}$ bears a positive value, the magnetic bond for NN parallel spins is shortened, and stretched for the NN antiparallel one, and vice versa. For simplicity, it is assumed that only Mn ions move, that is $d_{2i} \neq 0$ for Mn ions, $d_{2i-1} = 0$ for the immobile Co ions.

Combined with Eq. (2) and assumptions above, the Hamiltonian for the system can be further expressed as the following form

$$\mathcal{H} = -J_0^{\text{Co-Mn}} \sum_{i=1}^{N} \left( S_{2i-1} + S_{2i+1} \right) S_{2i} - J_2^{\text{Co-Co}} \sum_{i=1}^{N} S_{2i-1} S_{2i+1} - J_3^{\text{Mn-Mn}} \sum_{i=1}^{N} S_{2i} S_{2i+2}$$
$$- \mu_B g H \sum_{i=1}^{N} \left( S_{2i-1} + S_{2i} \right) - J_0^{\text{Co-Mn}} \eta \sum_{i=1}^{N} d_{2i} S_{2i-1} S_{2i} + J_0^{\text{Co-Mn}} \eta \sum_{i=1}^{N} d_{2i} S_{2i} S_{2i+1} \qquad (3)$$
$$- E \sum_{i=1}^{N} q d_{2i} + \frac{1}{2} k d_{2i}^2$$

Note that the sign of displacement $d_{2i}$ is different for neighboring unit blocks which are composed of NN Co and Mn ions, i.e., the alternating shrink and stretch between the NN bond will result in corresponding negative and positive variations of $d_{2i}$ in neighboring unit blocks.

To provide the exact solutions on this model, we divide the Hamiltonian of Eq. (1) into the following two parts

$$\mathcal{H}_{ps} = \frac{1}{2} k \sum_{i=1}^{N} \left[ d_{2i} - \frac{1}{k} \left( J_0^{\text{Co-Mn}} \eta S_{2i-1} S_{2i} - J_0^{\text{Co-Mn}} \eta S_{2i} S_{2i+1} + Eq \right) \right]^2 \qquad (4)$$

$$\mathcal{H}_s = -J_0^{\text{Co-Mn}} \sum_{i=1}^{N} \left( S_{2i-1} + S_{2i+1} \right) S_{2i} - J_2^{\text{Co-Co}} \sum_{i=1}^{N} S_{2i-1} S_{2i+1} - J_3^{\text{Mn-Mn}} \sum_{i=1}^{N} S_{2i} S_{2i+2}$$
$$- \mu_B g H \sum_{i=1}^{N} \left( S_{2i-1} + S_{2i} \right) - \frac{1}{2k} \left( J_0^{\text{Co-Mn}} \eta S_{2i-1} S_{2i} - J_0^{\text{Co-Mn}} \eta S_{2i} S_{2i+1} + Eq \right)^2 \qquad (5)$$

where $H_{ps}$ represents the Hamiltonian of one-dimensional linear harmonic oscillator, and its ground-state energy is only related with the angular frequency of natural oscillation.[22] Analogously, $\frac{1}{k} \left( J_0^{\text{Co-Mn}} \eta S_{2i-1} S_{2i} - J_0^{\text{Co-Mn}} \eta S_{2i} S_{2i+1} + Eq \right)$ corresponds to the displacement of oscillator away from the equilibrium position. The other part $\mathcal{H}_s$ represents the magnetic interaction energy which is only relevant to the coupling of spins. Now that the decoupling between the spins and phonons has been realized. It implies that the ground-state analysis can be accessible using $\mathcal{H}_s$ as a starting point, and the possibility of an implementation of the classical transfer-matrix method to gain exact results for thermodynamic quantities. By substituting the Hamiltonian $\mathcal{H}$ into the definition of the partition function, one obtains the following form

$$\mathcal{Z} = \text{Tr} \left( e^{-\beta \mathcal{H}} \right) = \text{Tr} e^{-\beta \left( \mathcal{H}_s + \mathcal{H}_{ps} \right)}$$
$$= \text{Tr}_s \text{Tr}_p \left( e^{-\beta \mathcal{H}_s} e^{-\beta \mathcal{H}_{ps}} \right) \qquad (6)$$
$$= \text{Tr}_s \left( e^{-\beta \mathcal{H}_s} \text{Tr}_p e^{-\beta \mathcal{H}_{ps}} \right) = Z_p Z_s$$

Here $\mathcal{Z}_p = \text{Tr}_p e^{-\beta \mathcal{H}_{ps}}$ and $\mathcal{Z}_s = \text{Tr}_s e^{-\beta \mathcal{H}_s}$ represent the partition functions related with linear harmonic oscillator and spin interactions respectively. $\beta = 1/(k_B T)$, $k_B$ is the Boltzmann's constant, $T$ is the absolute temperature. Although there exists spin symbol in $\mathcal{H}_{ps}$, it has no contribution to $\mathcal{Z}_p$, therefore the partition function $\mathcal{Z}$ can be expressed in the product of $\mathcal{Z}_p$ and $\mathcal{Z}_s$ in Eq. (6). It indicates that all the thermodynamics reflecting the main characters of magnetoelectric properties in Ca$_3$CoMnO$_6$ can be rigorously determined based on the partition function $\mathcal{Z}_s$.

By using unitary transformation, $\mathcal{Z}_s$ can be further simplified as

$$Z_s = \text{Tr}\left(e^{-\beta \mathcal{H}_s}\right) = \text{Tr}\left(\boldsymbol{V}^{-1}\boldsymbol{P}\boldsymbol{V}\boldsymbol{V}^{-1}\boldsymbol{P}\boldsymbol{V}\cdots\boldsymbol{V}^{-1}\boldsymbol{P}\boldsymbol{V}\right)$$
$$= \text{Tr}\,\boldsymbol{\lambda}^N = \lambda_1^N + \lambda_2^N + \lambda_3^N + \lambda_4^N \tag{7}$$

where $\boldsymbol{V}$ is a unitary matrix and satisfies the relation $\boldsymbol{V}^{-1}\boldsymbol{V} = \mathbf{I}$. $\boldsymbol{P}$ is the transfer-matrix with the matrix elements of

$$P_{(2i-1,2i),(2i+1,2i+2)} = \left\langle S_{2i-1}, S_{2i} \left| e^{-\beta \mathcal{H}^s_{(2i-1,2i)(2i+1,2i+2)}} \right| S_{2i+1}, S_{2i+2} \right\rangle \tag{8}$$

Apparently, the dimension of $\boldsymbol{P}$ is determined by the number of states that the subsystem possesses. Here, $S_i$ takes the value of $\pm 1$, therefore $\boldsymbol{P}$ is a $4 \times 4$ matrix with eigenvalues of $\lambda_1, \lambda_2, \lambda_3$ and $\lambda_4$. For clarity and convenience, $\lambda_1$ is assumed to be the largest eigenvalue in the following elaboration. $\mathcal{H}^s_{(2i-1,2i),(2i+1,2i+2)}$ stands for the cell Hamiltonian of $\mathcal{H}_s$, and involves all the magnetic interaction terms:

$$\mathcal{H}^s_{(2i-1,2i),(2i+1,2i+2)} = -J_0^{\text{Co-Mn}}\left(S_{2i-1} + S_{2i+1}\right)S_{2i} - J_2^{\text{Co-Co}}S_{2i-1}S_{2i+1} - J_3^{\text{Mn-Mn}}S_{2i}S_{2i+2}$$
$$- \frac{\mu_B g H}{2}\left(S_{2i-1} + S_{2i} + S_{2i+1} + S_{2i+2}\right)$$
$$- \frac{1}{2k}\left(J_0^{\text{Co-Mn}}\eta S_{2i-1}S_{2i} - J_0^{\text{Co-Mn}}\eta S_{2i}S_{2i+1} + Eq\right)^2 \tag{9}$$

If all the eigenvalues are nondegenerate and the largest one, $\lambda_1$, is real, in the thermodynamic limit the partition function is simply given by

$$\lim_{N \to \infty} \mathcal{Z}_s = \lambda_1^N \tag{10}$$

and the corresponding free energy per unit cell is obtained

$$f_s = -k_B T \ln \lambda_1 \tag{11}$$

According to quantum canonical ensemble theory, a thermodynamic average of an arbitrary quantity A is defined as $\left\langle A(T,H)\right\rangle = \dfrac{1}{Z}\text{Tr}\left(Ae^{-\mathcal{H}/(k_B T)}\right)$. Combining this definition and their concepts in statistical mechanics,[] the average magnetization, magnetic susceptibility, magnetic specific heat as well as electric polarization can be expressed as follow:

$$m = \frac{1}{N}g\mu_B\left\langle \sum_i S_i \right\rangle, \quad \chi_m = \frac{g^2\mu_B^2}{Nk_B T}\left(\left\langle \left(\sum_i S_i\right)^2 \right\rangle - \left\langle \sum_i S_i \right\rangle^2\right)$$

$$C_m = \frac{1}{Nk_BT^2}\left(\left\langle \mathcal{H}^2 \right\rangle - \left\langle \mathcal{H} \right\rangle^2\right), \quad P_c = \frac{1}{N}q\left\langle \sum_{i=1}^{N}d_i \right\rangle \tag{12}$$

By employing the transfer-matrix method, the final analytical form of the above quantities can be obtained in the same way. Considering that the detailed derivation process has been demonstrated in many literatures,[24,25] and thus for brevity it will not be reiterated here.

## III. Results and Discussion

In this section, we present our model calculation on the ground-state properties and thermodynamic behaviors, and qualitatively compare them with recent experimental results. The different role of the parameters will be analyzed to elucidate the nature of this multiferroic chain system. More importantly, our exact solutions will provide favorable evidence on the evolution of magnetic order in $Ca_3CoMnO_6$ detected by recent high-magnetic-field measurements.[20]

### A. Ground-State Properties

Characterization of the magnetic ground state in a diatomic Ising chain is crucial for revealing and confirming the origin of ferroelectricity in $Ca_3CoMnO_6$. And the specific effects induced by different parameters can also be accessible from the ground-state analysis. Bearing all this in mind, we adopt the Morita-Horiguchi technique[21] to construct all the possible spin patterns of the chain at zero temperature. With the consideration of translation invariance and global spin inversion symmetry, sixteen possible ground-state configurations with minimum energies are obtained as shown in Table I. Among these configurations, spin type II and type III both exhibit a twofold degeneracy, and their energy-equivalent spin patterns are indicated in brackets.

According to the results given in Table I, ground-state phase diagrams can be straightforwardly determined. Since the emerging of several parameters, a preferred method is to fix some parameters with reasonable values. In the light of the suggestions proposed by Yao et al. and Guo et al.,[13-15] the electric charge $q=2$ and elastic coefficient $k=200$ are adopted here. For the spin interacting, it is agreed that the NNN interaction is antiferromagnetic (AFM) coupling, while the NN interaction is still questioned. Although fits to magnetic neutron diffraction data suggest a FM interaction,[8] first-principle calculations[4,16] and mean-field theory[15] all support the AFM interaction. To illustrate the confusion from both experimental and theoretical works, calculations on these two cases are carried out, and a detail analysis is made. Before proceeding the discussion, we choose $|J_0^{\text{Co-Mn}}|$ as a reduced unit for convenience and introduce a set of dimensionless variables: $j_{\text{Co}} = J_2^{\text{Co-Co}}/|J_0^{\text{Co-Mn}}|$, $j_{\text{Mn}} = J_3^{\text{Mn-Mn}}/|J_0^{\text{Co-Mn}}|$, $t = k_BT/|J_0^{\text{Co-Mn}}|$, $k = k/|J_0^{\text{Co-Mn}}|$, $h = \mu_BgH/|J_0^{\text{Co-Mn}}|$, $E = Eq/|J_0^{\text{Co-Mn}}|$.

In Fig. 1, several typical ground state phase diagrams in the zero magnetic field are displayed. These profiles intuitively demonstrate the effects of $j_{\text{Co}}$, $j_{\text{Mn}}$ and $h_e$ on the ground-state spin orderings. For the NN spin pair with FM interaction presented in Fig. 1(a), when the two exchange parameters satisfy the condition, $\left| j_{Co} + j_{Mn} \right| \geq 1$, ground magnetic state manifests itself as $\uparrow\uparrow\downarrow\downarrow$ ($\uparrow\downarrow\downarrow\uparrow$) order, indicating that a weak antiferromagnetic interaction $j_{Co}$ only combines a strong interaction $j_{Mn}$, the $\uparrow\uparrow\downarrow\downarrow$ spin state generating ferroelectricity is guaranteed. This conclusion coincides with the parameters relationship suggested by Zhang et al obtained from density-functional calculation.[4] Nonetheless, these two possible magnetic configurations $\uparrow\uparrow\downarrow\downarrow$ and $\uparrow\downarrow\downarrow\uparrow$ yielding the opposite polarizations, always coexist in a same phase under this circumstance. As a result, the system dose not show the macroscopic electric polarization at arbitrary temperature. However, if a non-zero $h_e$ is applied (Fig. 1(c)), the situation will be remarkably changed. As is seen, ground-state degeneracy has been effectively relieved and $\uparrow\downarrow\downarrow\uparrow$ order is retained. Although the application of $h_e$ causes an extremely small difference on the phase boundaries, that is $j_{Mn} = -0.995 - j_{Co}$ between IV(VI) and I in Fig. 1(a), and $j_{Mn} = -0.994 - j_{Co}$ between VI and I in Fig. 1(c), but ensures the single domain of electric polarization, which all along has been ascribed to frustrated interchain interaction. This elevation effect of $h_e$ on magnetic orders rarely concerned in the past has been intuitively demonstrated in our work, hence inspiring us to deduce the experimental observation of ferroelectricity most possibly stems from dominated electric poling effect, but not the weak interchain interaction. Certainly, to fully account this issue, a more complex model with inclusion of interchain interactions should be considered, which is obviously beyond the current work. For the antiferromagnetic case presented in Fig. 1(b) and (d), the similar constraint relationship is exhibited, and the other spin pattern $\uparrow\uparrow\downarrow\downarrow$ is remained under the driven of external field $h_e$. One can also find when the system is subjected to $h_e$, $P_e$ at ground-state is unaffected by the coupling type between the NN spin pairs as the $\uparrow\downarrow\downarrow\uparrow$ and $\uparrow\uparrow\downarrow\downarrow$ orders yield the same magnitude and direction of $P_e$. Therefore to clarify the magnetic couplings issue, further calculations with inclusion of magnetic field $h$ effects are needed.

In Fig. 2 we plot typical ground state phase diagrams in the magnetic field $h$ versus exchange interaction $j_{Mn}$ plane. The data shows that the interaction for NN spin pairs either is FM or AFM, emergence of $P_e$ at ground-sate can always be realized, since both give rise to effective exchange striction, being consistent with the arguments made above. But comparing the evolution of field-induced magnetic order, it is not difficult to find the system driven by NN AFM interactions behaves in a way more close to the actual behaviors of $Ca_3CoMnO_6$.[20] However, the one with the FM interaction is easily fully polarized even at a moderate value of $h$, which obviously does not agree with experimental measurement. This field dependence of ground-state profile unambiguously and intuitively provides the strong evidence for the AFM arguments based on first-principles calculations and mean-field theory. Given this result, we focus our attention on the AFM case in the following discussion.

As is seen, the phase diagram for the AFM interaction has been shown to be rather rich, demonstrating a large variety of ground states. Two possible magnetization routines that this magnetoelectric-chain system may go through are presented under the prerequisite of ↑↑↓↓ ground-state order : (i) IV→II→I; and (ii) IV→VIII→II→I. And most encouragingly, these scenarios qualitatively reproduce the main features of two versions of magnetization profiles respectively detected in the experimental measurements, providing the necessary information for us to make a tentative exploration on the topic of debate, which is a real series of metamagnetic transition that occurs in Ca$_3$CoMnO$_6$ compound. For the former, it displays magnetic order evolutions from the zero-field ↑↑↓↓ spin pattern to the ↑↑↑↓ state with a broad magnetization plateau, and then to the saturation ↑↑↑↑ state. This magnetization process agrees well with the viewpoints expounded by Jo et al based on neutron-scattering data.[10] But if check the parameters of exchange couplings, one can find the proportion of $j_{Mn}$ to $j_{Co}$ is more than three times that this zero-temperature transitions can be realized. And the ratio becomes quite large with the decrease of $j_{Co}$. This scaling relationship is highly improper if one consults to earlier works.[4,15,16] It is therefore the actual magnetic-field induced spin-state in Ca$_3$CoMnO$_6$ is more complex and cannot fully explained by the magnetization routine exhibited in case (i), i.e., the data obtained from this elastic ANNNI model does not support the interpretations based on neutron powder diffraction in Ref. [10].

Recent Kim et al. measured magnetization in this Ising-chain magnet up to pulsed magnetic fields of 92 T. They found that two plateaulike steps appeared at 10T and 20T with values of ∼ 2.5$\mu_B$/f.u. and ∼4$\mu_B$/f.u. for $h$ applied along the chain direction $c$ axis. This result differs from that dc data in Ref. [10]. More importantly, they discovered that the final saturation was not ∼ 4$\mu_B$/f.u. at 20T, but 7.7$\mu_B$/f.u. at 85T.[20] These observations break the conventional recognition upon Ca$_3$CoMnO$_6$ and bring new insight into the magnetic structure, suggestive of multiple metastable states in this spin system. It is encouraging that this sequence of transitions is qualitatively generated by the elastic diatomic-Ising chain in a large parameter range, although the magnetic orders obtained here all belong to the collinear type. Hence we make an attempt to explain the evolved process detected by high-field measurement. At $h$=0, the system manifests itself as ↑↑↓↓ configuration, which has been accepted by researchers. As $h$ increases, our data displays an exotic ↑↓↑ state with 1/3 magnetization plateau marked by VIII, similar to second quasiplateau in pulsed-field data. It implies the possibilities of residual Ca$_3$CoMnO$_6$ insertions induced by nonstoichiometry in real crystalline materials. When $h$ increases to a moderate value, the third quasiplateau with ↑↑↑↓ state is achieved, corresponding to a fully polarized spin state of Co$^{2+}$ moments. Finally the system becomes completely ferromagnetic at higher $h$. Overall, these features are well consistent with the experimental observations given in Ref. [20]. And the scaling relationship among these parameters is accord with previous reports.[4] Our data reveals a different evolution of magnetic orders in $Ca_3CoMnO_6$ under the premise of Ising spins for both Co and Mn ions. However, it is worth mentioned the slope phenomenon appearing in the magnetization curve is not reproduced in our work, which has been attributed to canting of $Mn^{4+}$ by Kim et al. Hence a further refined model with inclusion of quasi-isotropic properties of Mn ions is necessary to illustrate the physical mechanism behind this phenomenon. For the sake of completeness, the boundary equations for every neighboring phases in Fig. 2(b) are summarized in Table II.

**B. Thermodynamic Behaviors**

In this section, we report on the behaviors of some relevant thermodynamic quantities on one hand to test the rationality of our model by comparing with experimental data; on the other hand to account for the difference and further illustrate the nature of magnetism-driven ferroelectricity. Our attention will be paid to the antiferromagnetic case with $-1.5 < j_{Mn} < -0.5$ , as the calculations for other cases do not agree with experimental measurement.

Figure 3(a) shows the field dependence of magnetization ($m$) per site for several distinct temperatures at $j_{Mn} = -1.1$ . As expected, two major features are observed in $m$ ($h$) for $t$=0 , a peculiar ↑↓↑ configuration with $m$=1/3 for $0.608 \leqslant h \leqslant 1.82$ and a broad 1/2 magnetization plateau for $1.82 \leqslant h \leqslant 4.2$, largely coinciding with those seen in Ref. [20]. And we notice that these zero-temperature transitions are very sensitive to the temperature, and smeared out by the thermal fluctuations even at a small value of $t$. Similar behaviors are exhibited in polarization ($P_c$) curve as well, see Fig. 3(b). At extreme low temperature, $P_c$ reaches its saturation and exhibits a plateaulike step. When $h$ increases close to critical field, $P_c$ rapidly decreases and completely collapses for $h$ beyond the critical value, which is just the boundary condition of ↑↑↓↓ → ↑↓↑ , confirming the intimate relation between ferroelectric order and magnetic order. This behavior is also quite well consistent with that experimental observation.[20] These good consistencies indicate the rationality of our model. It is particularly worth mentioning, $P_c$ receives evident suppression with temperature increasing for this sequence of zero-temperature transition, while in the case of $j_{Mn} < -1.5$ , the ability of $P_c$ against thermal fluctuations is more robust, overlapping at maximum in low-field region for $t$ up to 0.18 (not shown here). The contrast of $P_c$ responses in these two cases can be treated as an effective way to recognize the nature of magnetic structure in experiment.

To obtain a full comparison with experimental results, in Fig. 4 we plot the magnetic susceptibility $\chi_m$ and magnetic specific heat $C_m/t$ as a function of temperature at zero magnetic-field for various $j_{Mn}$ . As one can see, $\chi_m$ exhibits a round maximum upon temperature increasing, indicative of the short range of antiferromagnetic correlation. And its peak suffers suppression and shifts towards high temperature with the strengthen of antiferromagnetic coupling $j_{\mathrm{Mn}}$. The similar variation trend can be observed in specific heat curves as well. Although the main characters of $Ca_3CoMnO_6$ have been reproduced in our simulation, there are still some discrepancies upon the behaviors of $\chi_m$. The experimental data from Choi et al. shows that a weak ferromagnetism emerges at low-temperature in $\chi_m$ curve which is absent from our results. At first, we have thought this peculiar behavior stems from the application of external field. To verify this conjecture, we have calculated temperature dependence of magnetic susceptibility for several different magnetic fields. However, the weak ferromagnetism in $\chi_m$ is still not reproduced. It is therefore tentative for us to attribute this exotic phenomenon to the frustrated interchain interaction on the triangle lattice of *ab* plane, which suggests this simple Ising-chain model needed to be further refined.

## IV. Summary

In summary, an elastic diatomic-Ising chain model is adopted to describe the magnetoelectric properties of multiferroic $Ca_3CoMnO_6$ compound. By employing magneto-phono decoupling and Morita-Horiguchi technique, the ground-state configurations of this chain system has been rigorously determined. The computational data intuitively demonstrates the elevation effect of external electric field on the degenerated ground-state orders. And it confirms the parameter relationship satisfied for electric polarization emerging that the AF coupling between NNN spin pairs is needed to be close to or more than NN ones. Driven by the external magnetic field, the system presents a different sequence of zero-temperature transitions, displaying a peculiar $\uparrow\downarrow\uparrow$ state with 1/3 magnetization plateau, which provides the theoretical support for recent high-field measurement.

## IV. Acknowledgements

This work was supported by the National Natural Science Foundation of China (Grant Nos. 51102036 and 11474045) and Doctoral Starting up Foundation of Dalian Nationalities University, China.

[1] S. W. Cheong and M. Mostovoy, Nat. Mater. **6**, 13 (2007).

[2] W. Eerenstein, N. D. Mathur, and J. F. Scott, Nature **442**, 759 (2006).

[3] Y. Tokura, S. Seki, and N. Nagaosa, Rep. Prog. Phys. **77**, 076501 (2014).

[4] Y. Zhang, H. J. Xiang, and M.-H. Whangbo, Phys. Rev. B **79**, 054432 (2009).

[5] S. Picozzi, K. Yamauchi, B. Sanyal, I. A. Sergienko, and E. Dagotto, Phys. Rev. Lett. **99**, 227201 (2007).

[6] V. G. Zubkov, G. V. Bazuev, A. P. Tyutyunnik, and I. F. Berger, J. Solid State Chem. **160,** 293 (2001).

[7] S. Rayaprol, K. Sengupta, and E. V. Sampathkumaran, Solid State Commun. **128**, 79 (2003).

[8] Y. J. Choi, H. T. Yi, S. Lee, Q. Huang, V. Kiryukhin, and S.-W. Cheong, Phys. Rev. Lett. **100**, 047601 (2008).

[9] S. D. Kaushik, S. Rayaprol, J. Saha, N. Mohapatra, V. Siruguri, P. D. Babu, S. Patnaik, and E. V. Sampathkumaran, J. Appl. Phys. **108**, 084106 (2010).

[10] Y. J. Jo, S. Lee, E. S. Choi, H. T. Yi, W. Ratcliff, Y. J. Choi, V. Kiryukhin, S. W. Cheong, and L. Balicas, Phys. Rev. B **79**, 012407 (2009).

[11] H. Wu, T. Burnus, Z. Hu, C. Martin, A. Maignan, J. C. Cezar, A. Tanaka, N. B. Brookes, D. I. Khomskii, and L. H. Tjeng, Phys. Rev. Lett. **102**, 026404 (2009).

[12] T. Lancaster, S. J. Blundell, P. J. Baker, H. J. Lewtas, W. Hayes, F. L. Pratt, H. T. Yi, and S.-W. Cheong, Phys. Rev. B **80**, 020409(R) (2009).

[13] X. Y. Yao and V. C. Lo, J. Appl. Phys. **104**, 083919 (2008).

[14] X. Y. Yao, V. C. Lo, and J.-M. Liu, J. Appl. Phys. **106**, 013903 (2009).

[15] Y. J. Guo, S. Dong, K. F. Wang, and J.-M. Liu, Phys. Rev. B **79**, 245107 (2009).

[16] M. Nishida, F. Ishii, and M. Saito, JPS Conf. Proc. **3**, 014040 (2014); M. Nishida, F. Ishii, and M. Saito, J. Phys. Soc. Jpn. **83**, 124711 (2014).

[17] V. Kiryukhin, Seongsu Lee, W. RatcliffII, Q. Huang, H. T. Yi, Y. J. Choi, and S-W. Cheong, Phys. Rev. Lett. **102**, 187202 (2009).

[18] Z. W. Ouyang, N. M. Xia, Y. Y. Wu, S. S. Sheng, J. Chen, Z. C. Xia, and L. Li, Phys. Rev. B **84**, 054435 (2011).

[19] M. Y. Ruan, Z. W. Ouyang, S. S. Sheng, X. M. Shi, Y. M. Guo, J. J. Cheng, and Z. C. Xia, J. Magn. Magn. Mater. **344**, 55 (2013).

[20] J. W. Kim, Y. Kamiya, E. D. Mun, M. Jaime, N. Harrison, J. D. Thompson, V. Kiryukhin, H. T. Yi, Y. S. Oh, S.-W. Cheong, C. D. Batista, and V. S. Zapf, Phys. Rev. B **89**, 060404(R) (2014).

[21] T. Morita and T. Horiguchi, Phys. Lett. A **38**, 223 (1972).

[22] S. Weinberg, Lectures on Quantum Mechanics (Cambridge University Press, Cambridge, 2013).

[23] K. Stowe, An Introduction to Thermodynamics and Statistical Mechanics (Cambridge University Press, Cambridge, 2007).

[24] A. Du, H. J. Liu, and Y. Q. YǍu, Phys. Stat. Sol. (b) **241**, 175 (2004).

[25] A. Du, Y. Ma, and Z. H. Wu, J. Magn. Magn. Mater. **305**, 233 (2006).

TABLE I: The possible spin configurations at ground state and their energies per block.

| Label | Spin configuration | Energy per block |
|-------|-------------------|------------------|
| I | $\uparrow\uparrow\uparrow\uparrow$ | $\mathcal{H}_s = -2J_0^{\text{Co-Mn}} - J_2^{\text{Co-Co}} - J_3^{\text{Mn-Mn}} - \frac{1}{2k}(Eq)^2 - 2H$ |
| II | $\uparrow\uparrow\uparrow\downarrow(\uparrow\uparrow\uparrow)$ | $\mathcal{H}_s = -J_2^{\text{Co-Co}} + J_3^{\text{Mn-Mn}} - \frac{1}{2k}E^2q^2 - H$ |
| III | $\uparrow\uparrow\downarrow\uparrow(\downarrow\uparrow\uparrow\uparrow)$ | $\mathcal{H}_s = J_2^{\text{Co-Co}} - J_3^{\text{Mn-Mn}} - \frac{1}{2k}\left[E^2q^2 + \left(2J_0^{\text{Co-Mn}}\eta\right)^2\right] - H$ |
| IV | $\uparrow\uparrow\downarrow\downarrow(\downarrow\downarrow\uparrow\uparrow)$ | $\mathcal{H}_s = J_2^{\text{Co-Co}} + J_3^{\text{Mn-Mn}} - \frac{1}{2k}\left(Eq + 2J_0^{\text{Co-Mn}}\eta\right)^2$ |
| V | $\uparrow\downarrow\uparrow\downarrow(\uparrow\downarrow\uparrow\downarrow)$ | $\mathcal{H}_s = 2J_0^{\text{Co-Mn}} - J_2^{\text{Co-Co}} - J_3^{\text{Mn-Mn}} - \frac{1}{2}\frac{(Eq)^2}{k}$ |
| VI | $\uparrow\downarrow\downarrow\uparrow(\uparrow\uparrow\uparrow\downarrow)$ | $\mathcal{H}_s = J_2^{\text{Co-Co}} + J_3^{\text{Mn-Mn}} - \frac{1}{2k}\left(Eq - 2J_0^{\text{Co-Mn}}\eta\right)^2$ |
| VII | $\uparrow\uparrow\uparrow\downarrow\uparrow\uparrow$ | $\mathcal{H}_s = \frac{1}{3}\left\{-2J_0^{\text{Co-Mn}} + J_2^{\text{Co-Co}} + J_3^{\text{Mn-Mn}} - \frac{1}{2k}\left[3E^2q^2 - 8J_0^{\text{Co-Mn}}\eta Eq + 8\left(J_0^{\text{Co-Mn}}\eta\right)^2\right] - 2H\right\}$ |
| VIII | $\uparrow\uparrow\uparrow\uparrow\uparrow\uparrow$ | $\mathcal{H}_s = \frac{1}{3}\left\{2J_0^{\text{Co-Mn}} + J_2^{\text{Co-Co}} + J_3^{\text{Mn-Mn}} - \frac{1}{2k}\left[3E^2q^2 + 8\left(J_0^{\text{Co-Mn}}\eta\right)^2\right] - 2H\right\}$ |
| IX | $\uparrow\uparrow\uparrow\downarrow\downarrow\downarrow$ | $\mathcal{H}_s = \frac{1}{3}\left\{-2J_0^{\text{Co-Mn}} + J_2^{\text{Co-Co}} + J_3^{\text{Mn-Mn}} - \frac{1}{2k}\left[3E^2q^2 + 8\left(J_0^{\text{Co-Mn}}\eta\right)^2\right]\right\}$ |
| X | $\uparrow\uparrow\uparrow\uparrow\downarrow\downarrow$ | $\mathcal{H}_s = \frac{1}{3}\left\{2J_0^{\text{Co-Mn}} + J_2^{\text{Co-Co}} + J_3^{\text{Mn-Mn}} - \frac{1}{2k}\left[3E^2q^2 + 8J_0^{\text{Co-Mn}}\eta Eq + 8\left(J_0^{\text{Co-Mn}}\eta\right)^2\right]\right\}$ |
| XI | $\uparrow\downarrow\uparrow\uparrow\uparrow\downarrow\downarrow$ | $\mathcal{H}_s = J_0^{\text{Co-Mn}} - \frac{1}{2k}\left[2\left(J_0^{\text{Co-Mn}}\eta\right)^2 + 2J_0^{\text{Co-Mn}}\eta Eq + E^2q^2\right]$ |
| XII | $\uparrow\uparrow\uparrow\downarrow\downarrow\downarrow\uparrow$ | $\mathcal{H}_s = -J_0^{\text{Co-Mn}} - \frac{1}{2k}\left[2\left(J_0^{\text{Co-Mn}}\eta\right)^2 - 2J_0^{\text{Co-Mn}}\eta Eq + E^2q^2\right]$ |
| XIII | $\uparrow\uparrow\downarrow\downarrow\uparrow\downarrow\uparrow$ | $\mathcal{H}_s = J_2^{\text{Co-Co}} - \frac{1}{2k}\left[\left(2J_0^{\text{Co-Mn}}\eta\right)^2 + E^2q^2\right]$ |
| XIV | $\uparrow\uparrow\downarrow\downarrow\uparrow\uparrow\downarrow$ | $\mathcal{H}_s = J_3^{\text{Mn-Mn}} - \frac{1}{2k}\left[2\left(J_0^{\text{Co-Mn}}\eta\right)^2 + E^2q^2\right]$ |

TABLE II: Equations of the phase boundaries for $J_0^{\text{Co-Mn}} < 0$, $j_{\text{Co}} = -0.5$, and $h_e = 0.2$ (corresponding to the right panel Fig. 2(b)).

| Phase 1 | Phase 2 | Boundary |
|---------|---------|----------|
| IV | V | $j_{\text{Mn}} = -0.494$ |
| IV | II | $h = 1.012$ |
| IV | VIII | $h = -0.492 - j_{\text{Mn}}$ |
| V | VIII | $h = 0.99 + 2 j_{\text{Mn}}$ |
| VIII | III | $h = 0.99 - 4 j_{\text{Mn}}$ |
| VIII | II | $h = 4.02 + 2 j_{\text{Mn}}$ |
| II | I | $h = 2 - 2 j_{\text{Mn}}$ |
| III | I | $h = 3.01$ |

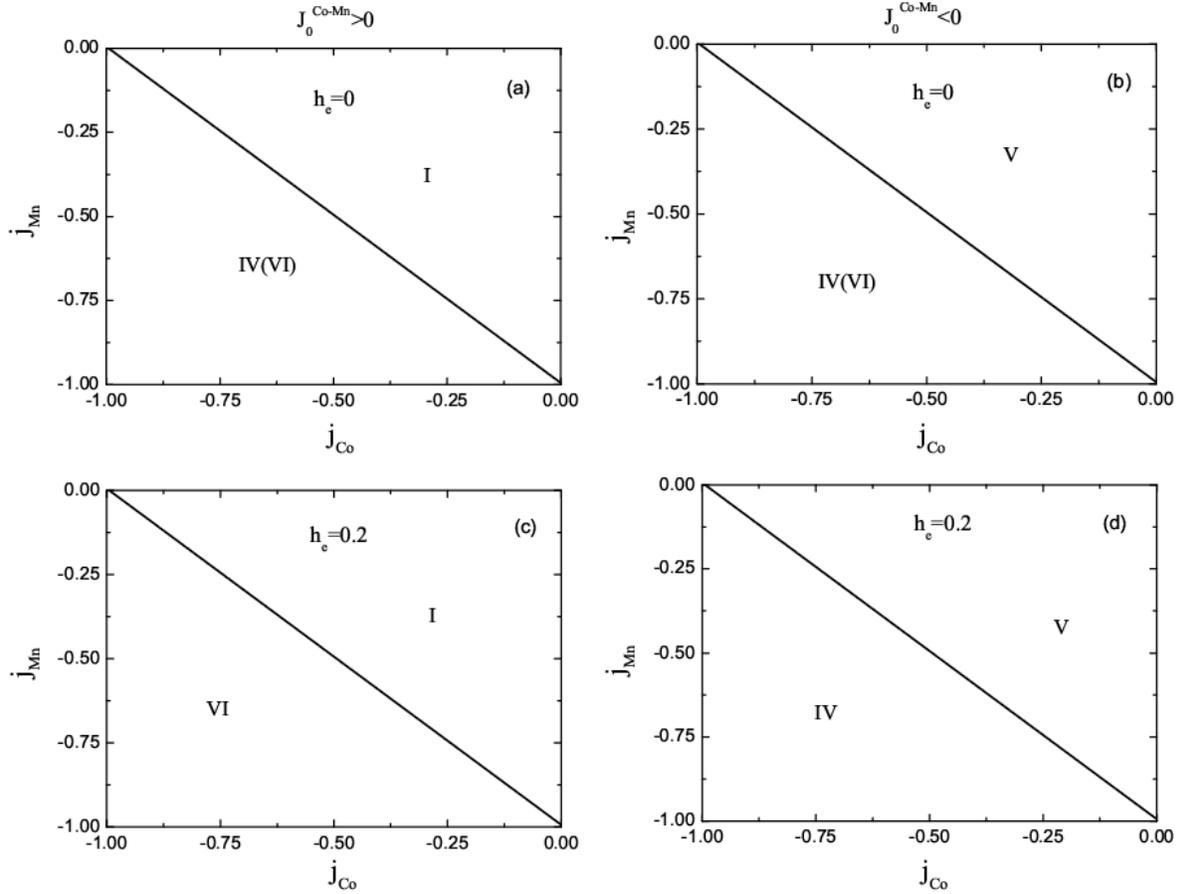

Fig. 1: Ground-state phase diagrams in the ($j_{Co}$, $j_{Mn}$) plane for $h$=0. Left panels (a) and (c) correspond to the ferromagnetic Heisenberg interaction $J_0^{Co-Mn} > 0$; right panels (b) and (d) to the antiferromagnetic one $J_0^{Co-Mn} < 0$. In upper panels (a) and (b) $h_e$=0 has been taken, while in lower panels (c) and (d) a non-zero electric field $h_e$=0.2 is applied along the chain.

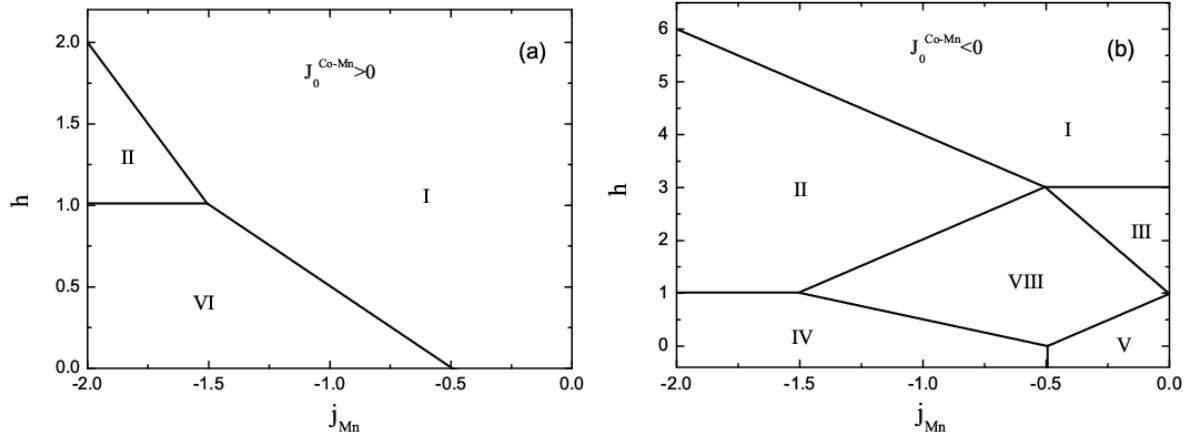

Fig. 2: Ground-state phase diagrams in the $j_{Mn}$-$h$ plane under $j_{Co} = -0.5$ and $h_e = 0.2$ for (a) ferromagnetic coupling $J_0^{Co-Mn} > 0$ and (b) antiferromagnetic coupling $J_0^{Co-Mn} < 0$.

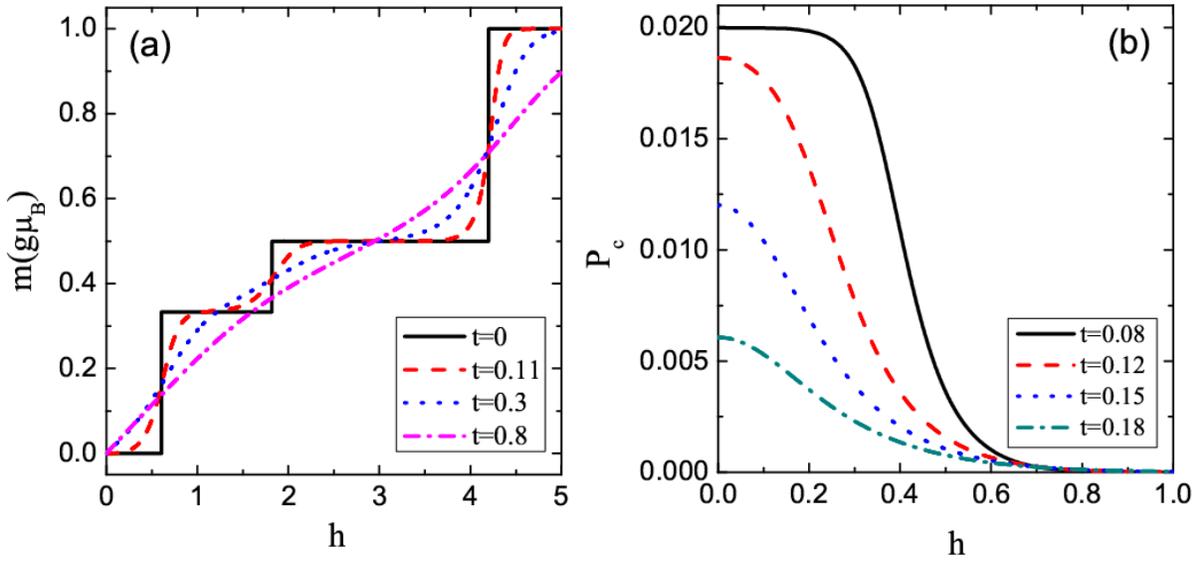

Fig. 3: Field dependence of (a) magnetization and (b) polarization for several typical temperatures and $j_{Co} = -0.5$, $j_{Mn} = -1.1$, $h_e = 0.2$.

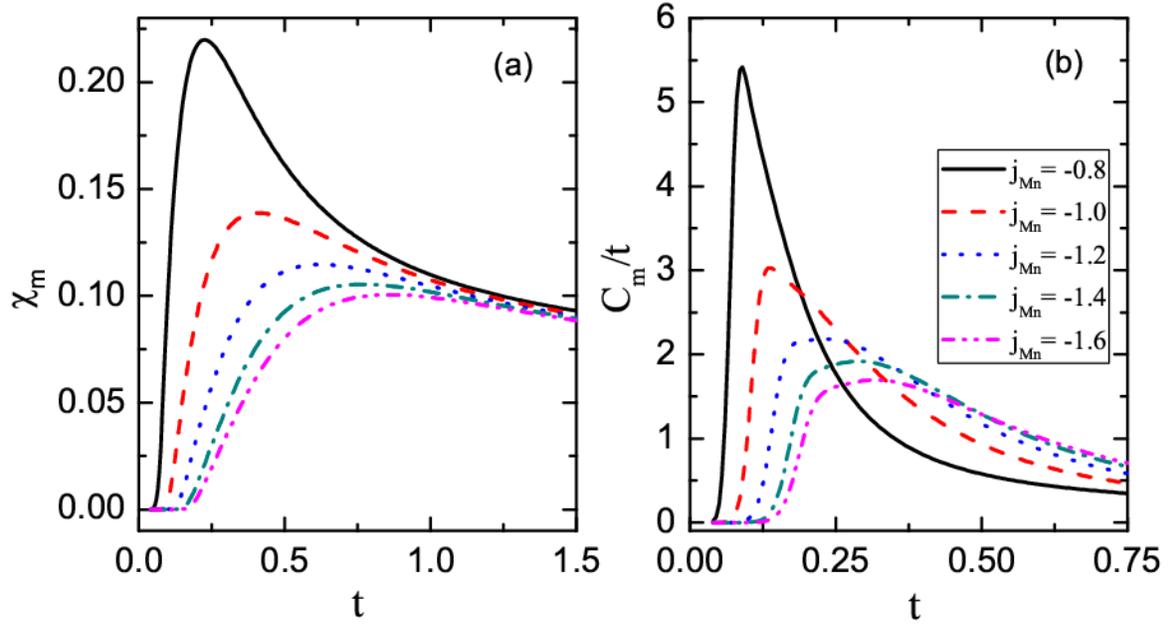

Fig. 4: (a) Magnetic susceptibility and (b) magnetic specific heat as a function of temperature at several typical NNN interactions between Mn ions for $j_{Co} = -0.5$, $h_e$=0.2 and $h$=0.